\documentclass{iau}
\usepackage{graphicx}

\title[Pulsations of blue supergiants] 
{Pulsations of blue supergiants before and after helium core ignition}

\author[Jakub Ostrowski \& Jadwiga Daszy\'nska-Daszkiewicz]   
{Jakub Ostrowski
 \and Jadwiga Daszy\'nska-Daszkiewicz}

\affiliation{Instytut Astronomiczny, Uniwersytet Wroc{\l}awski, ul. Kopernika 11, 51-622 Wroc{\l}aw, Poland\\
email: {\tt ostrowski@astro.uni.wroc.pl} \\[\affilskip]}

\pubyear{2014}
\volume{301}  
\pagerange{1--4}
\jname{Precision Asteroseismology}
\editors{J.A. Guzik, W.J. Chaplin, G. Handler \& A. Pigulski, eds.}
\begin{document}

\maketitle

\begin{abstract}
We present results of pulsation analyses of B-type supergiant models with masses of $14 - 18 M_\odot$, considering evolutionary stages before and after helium core ignition. Using a non-adiabatic pulsation code, we compute instability domains for low degree modes.
For selected models in these two evolutionary phases, we compare properties of pulsation modes. 
Significant differences are found in oscillation spectra and the kinetic energy density of pulsation modes.
\end{abstract}

\firstsection 
\section{Introduction}

Slowly Pulsating B-type supergiants (SPBsg) are a new class of pulsating variable stars. They have been discovered by \cite[Saio et al.~(2006)]{Saio06} who found 48 frequencies in the light variations of the blue supergiant HD 163899 (B2 Ib/II, \cite[Klare \& Neckel 1977]{KN77}, \cite[Schmidt \& Carruthers 1996]{SC96}) and attributed them to g- and p-mode pulsations. This was unexpected because it was believed that g modes cannot propagate in stars beyond the main sequence due to very strong radiative damping in the helium core.

The discovery has prompted a few groups (\cite[Godart et al.~2009]{Godart09}, \cite[Daszy\'nska-Daszkiewicz et al.~2013]{JDD13}) to reanalyse pulsation stability in models of B-type stars after the Terminal Age Main Sequence (TAMS) and to further studies of SPBsg variables. The presence of g-mode pulsations in B-type post-main-sequence stars has been explained by a partial reflection of some modes at an intermediate convective zone (ICZ) related to the hydrogen-burning shell or at a chemical gradient zone surrounding the radiative core. However, all studies of these objects published so far are based on the assumption that HD 163899 has not reached the phase of helium core ignition, i.e., it is in the phase of hydrogen shell burning. This assumption does not have to be made, because the blue loop can reach temperatures of early B spectral types.
In this paper we investigate this possibility and compare two SPBsg models: before and after He core ignition.

\section{Instability domains} \label{instab}

Our evolutionary models were calculated with the MESA evolution code (Modules for Experiments in Stellar Astrophysics, \cite[Paxton et al.~2011]{Paxton11}, \cite[Paxton et al.~2013]{Paxton13}). We adopted a hydrogen abundance at ZAMS of $X = 0.7$, metal abundance of $Z = 0.015$ and OPAL opacity tables (\cite[Iglesias \& Rogers 1996]{OPAL}) with the AGSS09 metal mixture (\cite[Asplund et al.~2009]{AGSS09}). We took into account convective overshooting from the hydrogen and helium core and inward overshooting from non-burning convective zones, using the exponential formula (\cite[Herwig 2000]{Herwig00}):
\begin{equation}
  \label{overshooting}
  D_\mathrm{OV} = D_\mathrm{conv}\exp(-\frac{2z}{f\lambda_P}),
\end{equation}
\noindent where $D_\mathrm{conv}$ is the mixing-length-theory-derived diffusion coefficient at a user-defined location near the core boundary, $\lambda_P$ is the pressure scale height at that location, $z$ is the distance in the radiative layer away from that location, and $f$ is an adjustable parameter, which we set to $0.01$. All effects of rotation and mass loss were neglected. We performed non-adiabatic pulsation analyses using the code of Dziembowski (\cite[1977]{Dz77}).

\begin{figure}
\begin{center}
 \includegraphics[width=4.8in]{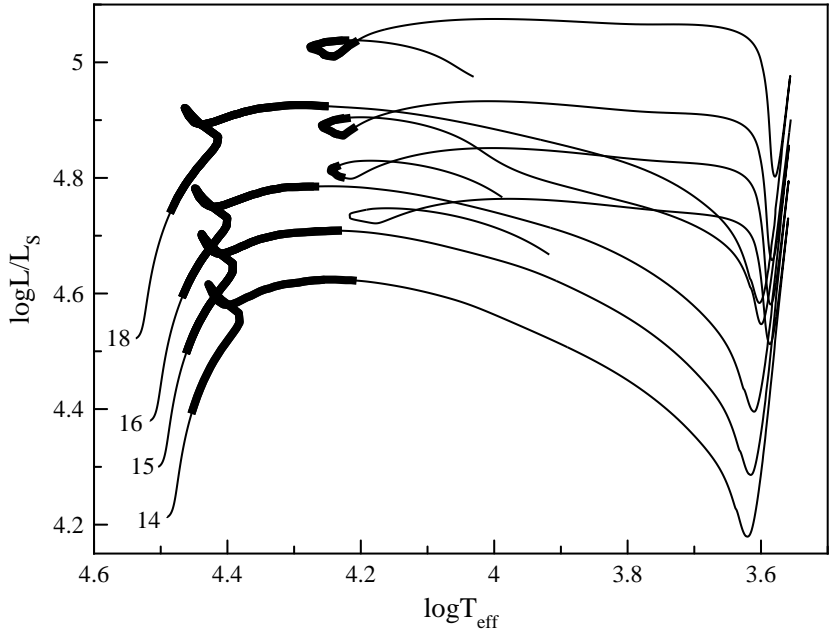}
 \caption{The H-R diagram with instability domains, marked as thick lines, for modes of degree $\ell = 0,1,2$ excited in models with masses of $14-18 M_\odot$.}
   \label{fig1}
\end{center}
\end{figure}

In Fig. \ref{fig1}, we present instability domains for the modes of the degree $\ell = 0,1,2$ excited in models with masses of $14-18 M_\odot$. There is an instability strip beyond the TAMS which is very similar to the one from previous calculations (e.g. \cite[Daszy\'nska-Daszkiewicz et al.~2013]{JDD13}). The main difference is the presence of unstable non-radial modes on the blue loops for more massive models ($M \gtrsim 14 M_\odot$ and $\log T_\mathrm{eff} \gtrsim 4.2$) whereas radial modes are stable in this evolutionary stage. The existence of pulsation instability on the blue loop, as well as the blue loops themselves, depend critically on the metallicity, $Z$. For example, for $Z = 0.02$ there are no unstable modes on the blue loops in the considered range of masses. More details will be given by Ostrowski \& Daszy\'nska-Daszkiewicz (in preparation).

\section{Pulsation modes before and after He core ignition} \label{energy}

We compare pulsation properties of two models with similar positions in the H-R diagram: one during the hydrogen shell burning phase ($16 M_\odot$, $\log T_\mathrm{eff} = 4.343$, $\log L/L_\odot = 4.705$, Model 1) and the second during core helium burning ($15 M_\odot$, $\log T_\mathrm{eff} = 4.244$, $\log L/L_\odot = 4.815$, Model 2). The main difference between these models is the presence of a convective core in the model on the blue loop. Beyond the core, the propagation diagrams of these two models are qualitatively very similar; in particular, both have a fully developed intermediate convective zone.

\begin{figure}
 \begin{center}
 \includegraphics[width=5.0in]{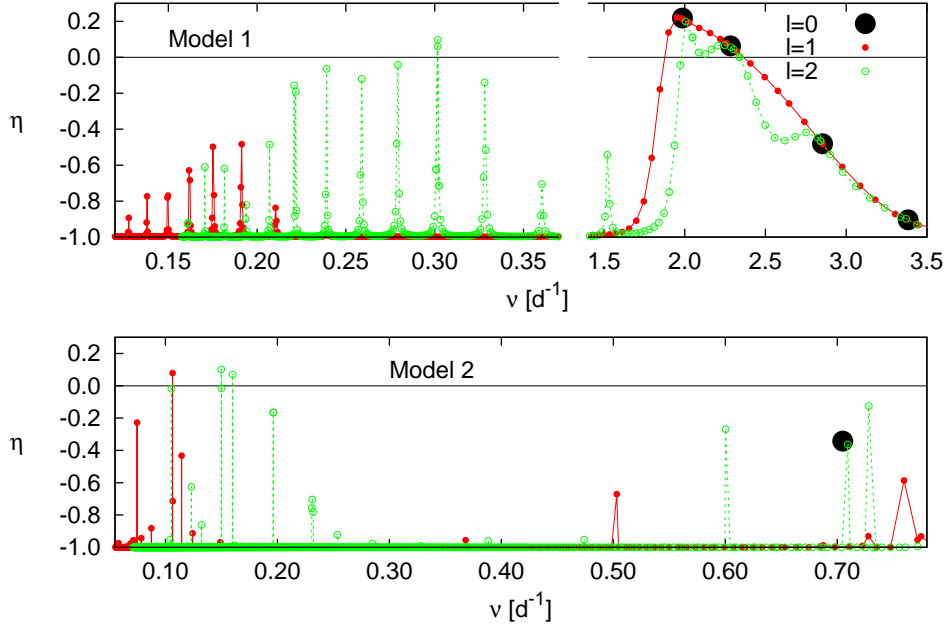}
 \caption{Instability parameter, $\eta$, for Model 1 (top panel) and Model 2 (bottom panel).}
 \label{fig2}
 \end{center}
\end{figure}

In Fig. \ref{fig2} we depict the instability parameter, $\eta$, as a function of frequency for Model 1 (top panel) and Model 2 (bottom panel). This parameter tells us whether the pulsation mode is unstable ($\eta > 0$) or not ($\eta \leq 0$). In the case of Model 1, we can see a regular structure controlled by mode trapping, and two instability regions can be identified (\cite[Daszy\'nska-Daszkiewicz et al.~2013]{JDD13}) whereas in Model 2 only a single very low frequency mode is excited.

\begin{figure}
 \begin{center}
 \includegraphics[width=5.4in]{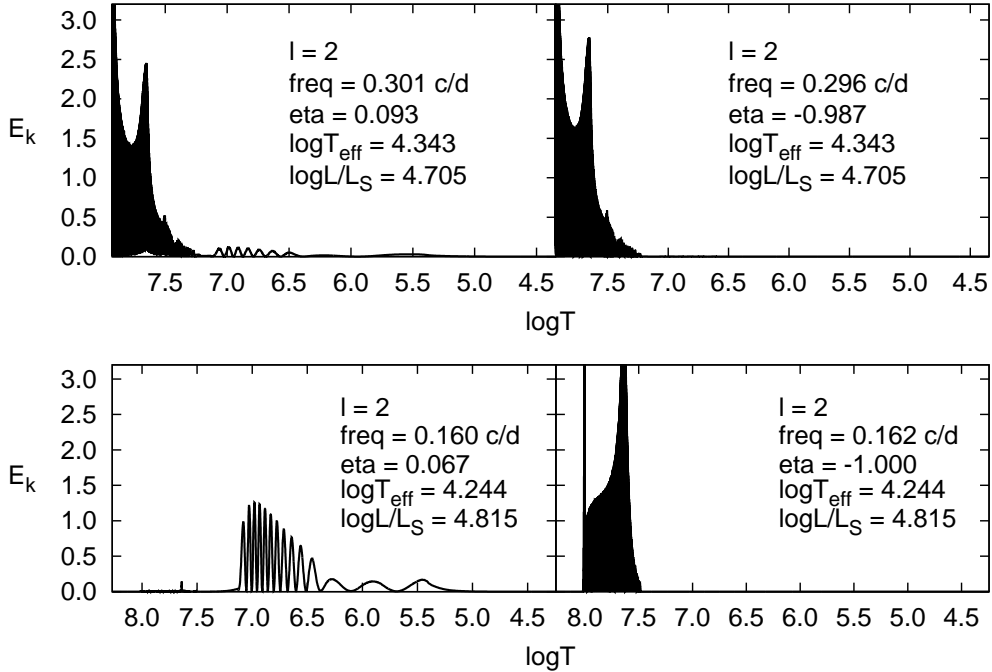}
 \caption{Top panel: the kinetic energy density of unstable (left panel) and stable (right panel) $\ell=2$ modes with close frequencies from Model 1. The values of $\nu$ and $\eta$ are listed in the panels. Bottom panel: the same as in the top panel, except for Model 2.}
 \label{fig3}
 \end{center}
\end{figure}

In the top panel of Fig. \ref{fig3}, we present the kinetic energy density for two very close frequency quadrupole high order g modes for the hydrogen shell-burning model. One of them (left panel) is unstable and the other stable (right panel). Their properties are described in the panels. For both modes, most of their energy is confined to the radiative helium core, where very strong damping occurs. The main difference between them is that due to a partial reflection at the ICZ, the unstable mode has some small amount of energy confined in the outer envelope. This energy is crucial for instability of pulsation modes in SBPsg stars and it is sufficient for the $\kappa$ mechanism operating in the Z-bump region to efficiently drive the mode.

A similar plot for the blue-loop model is shown in the bottom panel of Fig. \ref{fig3}. There are also two very close frequency quadrupole high order g modes, one unstable (left panel) and one stable (right panel). The behavior of $E_k$ is different for these modes than for the modes from the model before core helium burning. The entire energy of the stable mode is confined to the radiative zone in-between the convective core and the ICZ (there is strong damping in this area), whereas the unstable mode has almost all of its energy trapped in the outer radiative zone. That means that unstable modes on the blue loops have to be almost entirely reflected at the ICZ. This could explain why we observe in our models that undergo core helium burning many fewer unstable modes than in models before helium core ignition.

\section{Conclusions}

Our work has shown that blue loops can reach temperatures of B spectral types and there are unstable modes during this phase of evolution. It means that SPBsg stars might undergo core helium burning but whether they actually do or not is still an open question. We found that there is a huge difference in the behavior of the kinetic energy density of pulsation modes between models before and after helium core ignition. On the blue loop, the pulsation modes have to be almost entirely reflected at the ICZ in order to be unstable whereas for the models that undergo hydrogen shell-burning beyond the TAMS, a partial reflection at the ICZ is sufficient.

\end{document}